\documentclass[a4paper,11pt]{article}
\usepackage{pos}
\usepackage{slashed}
\usepackage{braket}
\usepackage{wrapfig}

\pagecolor{white}

\let\OLDthebibliography\thebibliography
\renewcommand\thebibliography[1]{
  \OLDthebibliography{#1}
  \setlength{\parskip}{0pt}
  \setlength{\itemsep}{0pt plus 0.3ex}
}

\title{The Proton EDM Experiment}

\author*[a]{Alexander Keshavarzi}

\affiliation[a]{Department of Physics and Astronomy, The University of Manchester, Manchester M13 9PL, U.K.}

\emailAdd{alexander.keshavarzi@manchester.ac.uk}

\abstract{The storage ring proton electric dipole moment (pEDM) experiment is the first direct search for a proton EDM. It will improve on the current (indirect) limit by at least $\mathcal{O}(10^4)$ and surpass the current sensitivity (set by neutron EDM experiments) to QCD CP-violation by at least $\mathcal{O}(10^3)$. This makes it an extremely sensitive probe of new physics, with major potential to discover the existence of axionic dark matter (and, therefore, solve the Strong CP problem) and find a new source of CP violation to help explain universe’s matter-antimatter asymmetry.}

\FullConference{42nd International Conference on High Energy Physics (ICHEP2024)\\
18-24 July 2024\\
Prague, Czech Republic\\}

\begin{document}
\maketitle

\section{Introduction and motivation}

The Standard Model (SM) of particle physics is an incomplete theory and cannot describe the observed universe. No source of dark matter has been found, no mechanism for charge-parity (CP) violation has been found to explain the overwhelming dominance of matter over antimatter in the universe,  and the Strong CP problem persists as a glaring, unsolved, fine-tuning problem in the SM's description of strong interactions via QCD.


Electric dipole moments (EDMs) of fundamental particles provide robust tests of the SM and are promising tools in the search for such physics beyond the SM (BSM). Crucially, they violate both time-reversal (T) and parity (P) symmetries which, assuming CPT symmetry, implies CP-violation. In general, particle EDMs have predicted values in the SM that are immeasurably small and therefore any experimentally measured, non-zero EDM signal would be clear evidence of BSM physics that would provide a new source of CP-violation.

A nucleon (proton or neutron) EDM can be generated from the CP violating phase in the CKM matrix through higher-order loop processes involving quark interactions. However, the suppression of these contributions at higher-order and the extremely small CP-violating phase of the CKM matrix mean nucleon EDMs in the SM are many orders of magnitude below current experimental sensitivity. A larger nucleon EDM can arise from BSM extensions of the SM which generate new CP-violating contributions to the CKM matrix, or due to the presence of the naturally arising $\bar{\theta}$-term in the QCD Lagrangian:
\begin{equation} \label{eq:LQCD}
\mathcal{L_{\rm QCD}} = -\frac{1}{4}G^a_{\mu\nu}G^{a\mu\nu} + \bar{q}(i\slashed{D} - m_q)q + \bar{\theta}\frac{g^2}{32\pi^2}G^a_{\mu\nu}\tilde{G}^{a\mu\nu} \, .
\end{equation}
Here, $G^a_{\mu\nu}$ is the gluon field strength tensor, $g$ is the strong coupling constant, $\tilde{G}^{a\mu\nu} = \frac{1}{2}\epsilon^{\mu\nu\alpha\beta}G^a_{\alpha\beta}$ and $\bar{q},q$ are the quark fields. The variable $\bar{\theta} = \theta + \varphi$, which is the sum of the QCD $\theta$ parameter and the quark mass phase $\varphi$, is the CP-violating parameter in QCD. However, no CP-violation has ever been observed in strong interactions, inexplicably implying that $\bar{\theta} \simeq 0$ due to an apparent, almost-exact cancellation of $\theta$ and $\varphi$. This is the Strong CP problem, and it remains one of the SM's most prominent fine-tuning problems to empirically explain the otherwise lack of observed CP-violation in the strong sector. This problem has a longstanding proposed solution in the Peccei-Quinn mechanism~\cite{Peccei:1977hh}, which gives rise to existence of a new particle and a common dark matter candidate, the axion. This solution satisfyingly removes the CP-violating $\bar{\theta}$-term from the QCD Lagrangian but, as-of-yet, the axion remains undiscovered. 


The presence of $\bar{\theta}$ in Eq.~\eqref{eq:LQCD} induces a non-zero nucleon EDM, $\vec{d}_N$, 
whose magnitude can be estimated via~\cite{CPEDM:2019nwp}
\begin{equation}
|d_N| \approx |\bar{\theta}|\frac{m_u m_d}{(m_u + m_d)} \frac{\mu_N}{\Lambda_{\rm QCD}} \approx |\bar{\theta}| \times 10^{-16} e\cdot {\rm cm} \, .
\end{equation}
Here, $m_u$ and $m_d$ are the masses of the up and down quarks, $\Lambda_{\rm QCD}$ is the QCD scale, $\mu_n = e/2m_N$ is the nuclear magneton, $e$ is the electric charge, and $m_N$ is the nucleon mass. 
The SM prediction for the nucleon EDM is $|d_N^{\rm SM}| \sim 10^{-31} e\cdot {\rm cm}$. The current best constraint on any $d_N$ comes from neutron (n) EDM experiments, with a direct experiment limit of $|d_n| < 1.8 \times 10^{-26} e \cdot {\rm cm}$~\cite{Abel:2020pzs}. The best constraint on the proton EDM of $|d_p^{\downarrow 199_{\rm Hg}}| < 2.0 \times 10^{-25} e \cdot {\rm cm}$ comes only indirectly from atomic physics experiments~\cite{Graner:2016ses}. 
\begin{wrapfigure}{R}{0.54\textwidth}
    \centering
   \includegraphics[width=0.53\textwidth]{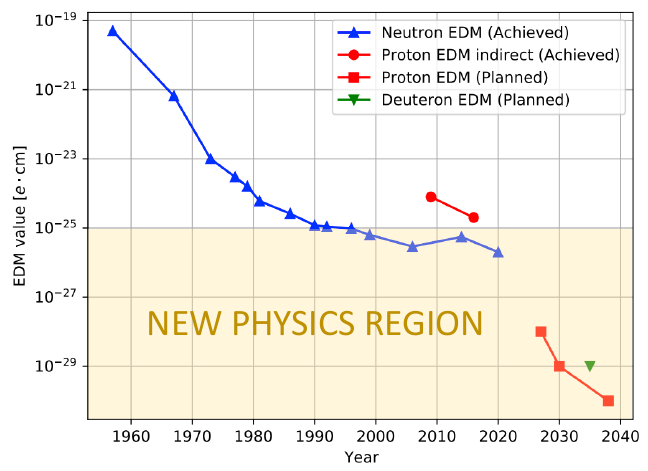}
   \caption{Projected sensitivity of the pEDM Experiment (red squares) compared to the current neutron EDM (blue triangles) and indirect proton EDM (red circles) limits. 
   The new physics range (Eq.~\eqref{eq:pEDM_NP}) is highlighted in yellow, emphasizing pEDM's power to cover most of this potential. This figure has been adapted from~\cite{Alexander:2022rmq}.}
   \label{fig:pEDM_limits}
   \vspace{0.1cm}
\end{wrapfigure}
Therefore, any confirmed direct measurement of a static nucleon EDM with a value in the range
\begin{equation} \label{eq:pEDM_NP}
10^{-26} e\cdot {\rm cm} \gtrsim |d_N| \gtrsim 10^{30} e\cdot {\rm cm} \, , 
\end{equation}
would provide a new source of CP-violation arising from either a value of the QCD $|\bar{\theta}|$ term in the corresponding range $10^{-10} \gtrsim |\bar{\theta}| \gtrsim 10^{14}$, or unambiguous BSM physics.  

The pEDM Experiment will be the first direct search for the proton EDM and will achieve a new paradigm of precision in nucleon EDMs by reaching $\lesssim 10^{-29} e\cdot{\rm}cm$, improving on the current limit by at least $\mathcal{O}(10^4)$. This will improve upon the current sensitivity to $|\bar{\theta}|$ by at least $\mathcal{O}(10^3)$. Crucially, and as indicated in Fig.~\ref{fig:pEDM_limits}, it will cover the majority of the unexplored range for new physics given in Eq.~\eqref{eq:pEDM_NP} and far surpass all other EDM measurement efforts in getting close to its SM prediction. Recent theoretical studies have additionally suggested that both the magnitude of $d_N$ and its sensitivity to new physics are charge-dependent, and that searches for the proton EDM are a potentially more sensitive probe of BSM physics than in the case of neutron EDM searches~\cite{DiLuzio:2020wdo,Smith:2023htu,DiLuzio:2023ifp}.

The proposed pEDM Experiment can measure a wide range of time-variations of the proton EDM. This is particularly important due to the oscillating behavior of the axion (or axion-like-particle) field, $a(t)$, which has been hypothesized to explain the dark matter content in the universe. Assuming the QCD axion couples to gluons and a total energy density of the axion field equal to the observed dark matter density, a time-dependent, oscillating proton EDM $d_p(t)$ has the form~\cite{CPEDM:2019nwp}:
\begin{equation}
d_p(t) \approx \frac{a(t)}{f_a} \times 10^{16} \approx 5 \times 10^{-35} \cos\left(\frac{1}{\hbar}m_ac^2(t-t_0) + \phi_0\right) \,e\cdot {\rm cm} \, .
\end{equation}
Here, $f_a$ is axion decay constant, $|t-t_0|$ is the measurement period and $\phi_0$ is some undetermined local phase. Such time variation is detectable by the pEDM Experiment, and a confirmed measurement which would be direct evidence of axionic dark matter and would provide a solution to the Strong CP problem. pEDM is sensitive to axion frequencies in the range $1~{\rm mHz} \lesssim \omega_a \lesssim 1~{\rm MHz}$, and axion masses in the range $10^{-7}~{\rm eV} \gtrsim m_a \gtrsim 10^{-22}~{\rm eV}$~\cite{CPEDM:2019nwp}. Importantly, recent theoretical work has suggested that an oscillating pEDM signature could be up to $\mathcal{O}(10^2)$ more prominent for the proton EDM than for the neutron~\cite{DiLuzio:2020wdo,Smith:2023htu,DiLuzio:2023ifp}.

Potential BSM contributions to the proton EDM are loop-induced from higher-order effects. As such, the potential new physics reach of the pEDM Experiment covers a wide range of interactions and energy scale. Assuming a NP scale $\Lambda_{\rm NP} \sim 1$ GeV and small couplings, e.g. $g \lesssim 10^{-5}$, the search for a proton EDM will probe light, weak new physics complimentary to searches at e.g.~LZ, LDMX, FASER and SHiP. For $\Lambda_{\rm NP} \gtrsim 1$ TeV, the NP contribution to the proton EDM can be estimated via~\cite{CPEDM:2019nwp}
\begin{equation}
|d_p^{\rm NP}| \approx \frac{g^2}{16\pi^2}\frac{e m_q}{\Lambda_{\rm NP}^2}\sin{\phi^{\rm NP}} e\cdot {\rm cm} \, ,
\end{equation}
where $m_q$ is the mass of a one-loop quark contribution and $\phi^{\rm NP}$ is a complex CP-violating phase. In this case, the new physics sensitivity can reach $\Lambda_{\rm NP} \sim 10^{3}$ TeV, making pEDM complimentary to searches for heavy new physics at the LHC, and proposed FCC and muon collider programs.

\section{Experimental methodology}

EDMs are a measure of a charge distribution of the particle being studied. A non-zero proton EDM would imply an uneven separation of the charges within the proton, and the presence of an external electric field on this uneven charge would result in a vertical tilt in the proton's polarization. Therefore, measuring a proton EDM requires storing longitudinally polarized protons in an electric field and, after a given time, using a polarimeter to detect the polarization angle of the proton. A signal of a vertical polarization component would be indicative of a static, non-zero proton EDM.

\begin{wrapfigure}{T}{0.54\textwidth}
\vspace{-1.2cm}
\begin{center}
\includegraphics[width=0.53\textwidth]{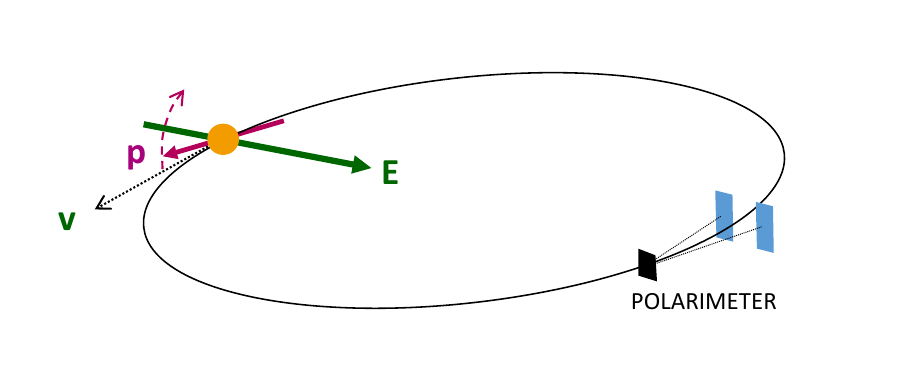}
\end{center}
\vspace{-0.5cm}
\caption{Diagram of the pEDM experimental concept and storage ring, with the horizontal spin precession locked to the momentum precession rate (“frozen” spin). The radial electric field acts on the particle EDM for the duration of the storage time. Two downstream polarimeters, one left and one right of the beam scattering target, provide the ability to measure the left-right asymmetry of the proton-carbon scattering over the beam storage time, where a measurable asymmetry is indicative of a static, non-zero EDM. This figure is taken directly from~\cite{Alexander:2022rmq}.}\label{fig:pEDM}
\vspace{-0.2cm}
\end{wrapfigure}
In the case of the pEDM experiment, the approach to measure this phenomenon requires a beam of $\sim 2\times 10^{10}$ highly polarized (spin-aligned) protons to be stored in a highly symmetric storage ring using electric field bending for $10^3$s (half the proton spin-decoherence time of $\sim 2\times 10^{3}$s). As shown in Fig.~\ref{fig:pEDM}, the electric field acts along the radial direction toward the center of the ring and will induce a vertical tilt in the polarization of the stored protons if an EDM exists. More specifically, as the electric field is perpendicular to the spin axis of the proton (and therefore perpendicular to the axis of the EDM), the proton's spin will precess in the vertical plane. With the polarization of injected protons known, high-precision measurements of the polarization of the protons after storage are then the essential deliverable, where the appearance of a vertical polarization component with time is the signal for a non-vanishing EDM.

When charged particles are placed in magnetic (B) and electric (E) fields, their spin precesses due to their magnetic dipole moment (MDM) and EDM, respectively, and these quantities are entangled. For a particle with mass $m$, charge $q$, velocity $\beta$, and anomalous magnetic moment $a=(g-2)/2$, the corresponding spin precession vector is given by:
\begin{equation} \label{eq:TBMT}
\vec{\omega} = \vec{\omega}_{\rm MDM} +  \vec{\omega}_{\rm EDM}  \approx \frac{e}{m} \left( \, \left[a\vec{B} + \left(a - \frac{1}{\gamma^2-1}\right)\left(\vec{\beta}\times\vec{E}\right)\right]_{\rm MDM} + \left[\frac{\eta}{2}\left(\frac{\vec{E}}{c} + \vec{\beta}\times\vec{B}\right) \right]_{\rm EDM} \, \right)\, .
\end{equation}
Here, $\gamma$ is the Lorentz factor and $\eta$ describes the coupling of the EDM of the particle to its spin via $\vec{d} = \eta\frac{q\hbar}{2mc}\vec{S}$. Measuring the spin properties of stored particles in this way is the basis of the highly successful Muon $g-2$ Experiment at Fermilab that has achieved an unprecedented 200 parts-per-billion precision on the muon's anomalous magnetic moment, $a_\mu$~\cite{Muong-2:2024hpx}. This has been possible by exploiting Eq.~\eqref{eq:TBMT} and overcoming the need for precise knowledge of the E-field by choosing a specific momentum for the stored muons - the ``magic'' momentum, $a_\mu-1/(\gamma_{\mu,{\rm magic}}^2 - 1) = 0$ - which results in the cancellation of the dominant electric field contribution to first order. Then, by precisely measuring the B-field, $\vec{\omega}_{\rm MDM}$ and  $\vec{\omega}_{\rm EDM}$ can then be individually extracted. However, the muon EDM's precision is limited at the Muon $g-2$ Experiment by the dominant MDM signal and the precision by which the B-field is known. 

The pEDM Experiment will achieve significant sensitivity to the proton EDM by building on the successes of the Muon $g-2$ Experiment and again exploiting Eq.~\eqref{eq:TBMT} to implement a novel experimental method: the ``frozen-spin'' technique. Under this scheme, the protons are chosen to have a different specific momentum, $p_{\rm magic} = 0.7007$ GeV/c, which results in 
\begin{equation}
a_p\vec{B} + \left(a_p - \frac{1}{\gamma^2-1}\right)\left(\vec{\beta}\times\vec{E}\right) = 0 \, ,
\end{equation}
thereby ``freezing'' the dominant spin precession component from the proton MDM and leaving it defined purely by the EDM component (to first order) and the magnitude of the E-field. Using this technique is a principle aim of pEDM and will result in the $>\mathcal{O}(10^4)$ improvement in sensitivity. 

With the MDM component nullified and the polarization of the stored protons at injection known, the appearance of a vertical polarization component with time is the signal for a non-vanishing EDM. The beam of protons will be continuously scattered off a carbon target and two downstream polarimeters, one left and one right of the target, measure the left-right (L-R) scattering asymmetry over the storage time (see Fig.~\ref{fig:pEDM}). The vertical polarization component of the protons, $p_y \propto (L-R)/(L+R)$, is proportional to the asymmetry in the L-R rates of the scattered beam. A measurable asymmetry is indicative of a non-zero EDM. 

The use of a highly symmetric, hybrid electric-magnetic storage ring (using electric bending and magnetic focusing) allows for simultaneous clockwise (CW) and counterclockwise (CCW) beam storage~\cite{Alexander:2022rmq}. This will facilitate significant cancellations of dominant, time-conserving sources of systematic uncertainties that mimic the time-violating signal of the EDM by inducing an apparent tilt in the proton's vertical polarization. By measuring both CW and CCW beams in the enhanced ring-lattice symmetry, such time-symmetric systematic effects acts as the time-reverse of each other, resulting in a cancellation of the effect and leaving the measured signal defined purely by the time-violating EDM. 

\section{Progress and future plans}

Major work has already been completed to move pEDM towards a functioning experiment. The experimental design, engineering and modelling of the experiment are complete. Mechanical prototypes for e.g. electric field bending are under construction, funded by a US Department of Energy Laboratory Directed R\&D grant. The measurement techniques are well understood, and the key systematic effects have been simulated and quantified. 

Upcoming work to move the experiment to technical design report (TDR) phase include e.g.~precision studies of the stored beam with realistic simulation of the beam over the necessary storage time ($10^3$ protons for $10^3$ s) and development of prototype silicon-tracking polarimeters~\cite{Gooding_2022}. 
\begin{wrapfigure}{r}{0.71\textwidth}
\vspace{-0.5cm}
\begin{center}
\includegraphics[width=0.7\textwidth]{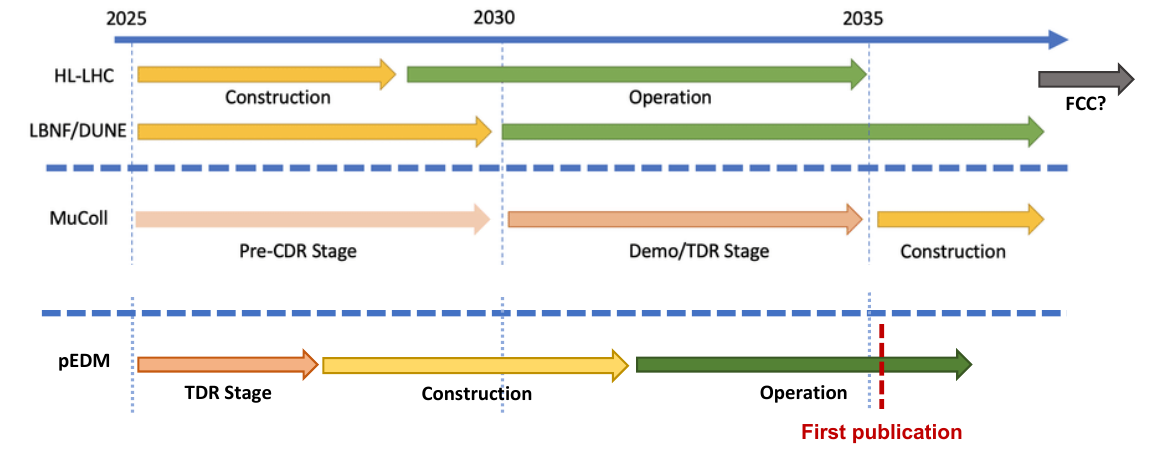}
\end{center}
\vspace{-0.5cm}
\caption{Estimated timeline the pEDM Experiment compared to other experimental particle physics programs. This figure has been adapted from~\cite{Black:2022cth}.}\label{fig:timeline}
\vspace{-0.4cm}
\end{wrapfigure}
Much like the high return of the Muon $g-2$ Experiment over a short timescale, it is expected that pEDM can go from TDR phase to final publication in $\sim20$ years.
 An estimated 10+ year timeline for pEDM compared to other future programs is shown in Fig.~\ref{fig:timeline}.

\section{Conclusions}

At the level of proposed precision, the pEDM Experiment is one of the best hopes for discovering a new source of CP violation to help explain the universe's matter-antimatter asymmetry and has a BSM sensitivity ranging from 1 GeV - 1 PeV. It is the first direct search for the proton EDM and will build upon the successful techniques of the Muon $g-2$ Experiment to improve on the current indirect limit by at least $\mathcal{O}(10^4)$, taking it close to the SM prediction of the proton EDM. A measured oscillating proton EDM signal would be evidence for the existence of axionic dark matter and therefore has the potential to provide a solution to the Strong CP problem. The experimental effort is moving towards TDR phase and is expected to reach final publication in $\sim20$ years, providing critical information to inform decisions regarding future particle physics programs.

\section*{Acknowledgements}
I would like to thank Luca Di Luzio for numerous useful discussions. This work is supported by The Royal Society (URF$\backslash$R1$\backslash$231503).

\end{document}